\documentstyle[aps,prl,epsf,floats]{revtex}
\begin{document}


\draft

\twocolumn[\hsize\textwidth\columnwidth\hsize\csname @twocolumnfalse\endcsname

\title{Boundary effects in reaction-diffusion processes\\}

\author{M. J. E. Richardson and Y. Kafri}

\address{Department of Physics of Complex Systems, The Weizmann Institute 
        of Science, Rehovot 76100, Israel\\[-2mm] $ $}
        
\maketitle
\vspace{-0.5cm}
{\center Date: \today. \par}

\begin{abstract}
The effects of a boundary on reaction systems are examined in the framework of the general single-species reaction/coalescence process. The boundary naturally represents the reactants' container, but is applicable to exciton dynamics in a doped TMMC crystal. We show that a density excess, which extends into the system diffusively from the boundary, is formed in two dimensions and below. This implies a surprising result for the magnetisation near a fixed spin in the coarsening of the one-dimensional critical Ising model. The universal, dimensionally-dependent functional forms of this density excess are given by an exact solution and the field-theoretic renormalisation group.

\end{abstract}
\vspace{2mm}
]


Reaction systems have been given a good deal of attention in the literature over the last few decades. They are widespread in nature and methods developed in their analysis have an applicability that extends well beyond conventional chemical systems. Moreover, they provide excellent examples of dynamical, many-body statistical processes and can exhibit a variety of interesting effects such as spontaneous symmetry breaking \cite{SYMBREAK} and pattern formation \cite{RACZ}.

In this work, we show that the imposition of impenetrable boundaries on a reaction system leads to a non-trivial spatial variation in the reactant density. Even though such a boundary would naturally represent the vessel holding the reactants, until now no such studies have appeared in the literature. In systems with many interacting degrees of freedom, boundaries often give rise to surface effects that penetrate far into the bulk. Our results demonstrate that in reaction systems such long-range effects are indeed present and have a high degree of universality.

We choose the class \cite{KANGRED,PELITI} of the general single-species reaction processes as the starting point for the study of these boundary effects in reaction systems, as it will provide a basis for the analysis of more complex systems. This universality class, comprising the annihilating random walk $A+A\rightarrow O$, coalescing random walk $A+A\rightarrow A$ and any combination thereof, is fundamental in the theoretical study of reaction systems and covers a broad range of physical phenomena. For example, the coalescence process is seen in the dynamics of excitonic annihilation reactions in the TMMC crystal \cite{TMMC,PRIV}. The predicted decay exponent from theory is in agreement with experiment for over five orders of magnitude. It should be noted that the boundary effects we introduce here are exhibited in the TMMC crystal in the presence of Mg$^{2+}$ or Cd$^{2+}$ doping. These defect ions act as perfect reflectors for the annihilating excitons \cite{MGCD}. 

A mapping also exists from the annihilating random walk to the domain coarsening dynamics in the critical one-dimensional Ising magnet \cite{GLAUBER,ISING,BRAY}. In this mapping the $A$ particles represent domain walls with an annihilation of two $A$ particles corresponding to a domain shrinking to zero size in a background of the opposite phase. An impenetrable boundary in the reaction system corresponds to a fixed boundary spin in the Ising magnet. Our analysis will show that domain walls are more likely to be found near the fixed spin than far away. This gives the counter-intuitive result that the absolute value of the dynamic, coarse-grained magnetisation is actually {\it lower} near a fixed spin.

In the interests of notational simplicity, we provide here the analysis specifically for the process $A+A\rightarrow O$. However, all densities given can be converted to the result for the coalescence process $A+A\rightarrow A$ by a simple factor of two. The annihilating random walk has been studied extensively in homogeneous unbounded systems, either of infinite extent or with periodic boundary conditions. Throughout the work, this will be referred to as the {\it bulk} case. It is well known \cite{KANGRED,PELITI,BPL,LUSH} that the variation of the density $\varrho$ with time $t$ differs from the mean-field prediction $\varrho\propto 1/t$ for dimensions $d\leq2$. In fact the actual density decays are $\varrho\propto t^{-1}\log t$ ($d=2$), and $\varrho\propto t^{-d/2}$ ($d<2$) with a universal amplitude. In all cases, it must be stressed that the density remains uniform throughout the system.

In the following, we will first introduce the boundary into the reaction process and specify the model to be studied. The mean-field approximation will be shown to predict a homogeneous density unchanged from the bulk case. However, an argument will be presented to show that the mean-field prediction breaks down in low dimensions. The central result of this work is that a {\it fluctuation-induced} excess density develops at the boundary and extends into the system diffusively. We outline the field-theoretic renormalisation group (RG) description which we use to identify the universal quantities of this excess. These calculations were performed in real space to one-loop order and show that in two dimension and below, the density excess $\varrho_E$ has the following form
\begin{eqnarray}
\varrho_E&=&\frac{1}{(8\pi Dt)^{d/2}}f_d\left(\frac{z^2}{2Dt}\right).  
\end{eqnarray}
Here, $D$ is the diffusion constant of the reactants, $z$ is the normal distance from the boundary and $f_d$ is the dimensionally dependent scaling function. We were able to find the late-time scaling functions $f_2$ and $f_1$ exactly in both two and one dimensions (given in Eqs. (\ref{surf}) and (\ref{1dexact}) respectively). The former we derive from the RG improved field-theoretic calculation and the latter from an exact solution. Finally, in the context of surface critical phenomena, the behaviour in the related {\it ballistic annihilation} reaction system will be briefly examined and compared with the diffusive case. 

We now introduce the model. The system is defined on a hypercubic, $d$-dimensional lattice with a lattice spacing of unity. The lattice, is infinite in $d-1$ transverse dimensions and semi-infinite (sites $1,2,\cdots,\infty$) in what will be called the $z$ direction. At time $t=0$ the lattice is filled with an initial density $\varrho_0$ of identical particles that perform two types of dynamics: diffusion and mutual annihilation.  The diffusion is represented by each particle hopping at a rate $D$ to any neighbouring site at random. A hop from site $z\!=\!1$ towards the boundary is disallowed. The diffusion is independent for each particle, and hence multiple occupancy of each site is possible, leading to bosonic particle statistics. However, if there are $n\geq2$ particles on a particular site, a reaction can occur there with a rate $\lambda n(n-1)$ to reduce $n$ by 2.

The above dynamics can be approximated by a {\it mean-field} description. This involves ignoring all possible correlations by considering a self-consistent equation involving just the average density $\overline{\varrho}$. In the continuum limit the boundary is enforced by a zero-current restriction, thus
\begin{eqnarray}
\partial_t\overline{\varrho}=D\nabla^2\overline{\varrho}-2\lambda\overline{\varrho}^2\;\;&\mbox{ with }&\;\;\partial_z\overline{\varrho}|_{z=0}=0. 
\end{eqnarray}
The boundary restriction is compatible with the bulk solution $\overline{\varrho}=\varrho_0/(1+\varrho_0 2\lambda t)$. Hence, in the absence of strong fluctuations, the density is uniform throughout the system. 

However, correlations must be properly accounted for in low dimensions. Consider the dynamics of the model, in one dimension, up to a time $t$ and far from the wall. Because random walks in one-dimension are recurrent, most particles within a diffusion length $l_b\sim\sqrt{2Dt}$ in the bulk, will have interacted and annihilated. This leads to a density in the bulk of the system of $\varrho_b\simeq l_b^{-1}\simeq c_b/\sqrt{t}$. However, close to the wall the diffusion length is smaller, $\varrho_w\simeq l_w^{-1}\simeq c_w/\sqrt{t}$ since $c_w\!>\!c_b$, leading to a density excess near the boundary.

Nevertheless, this argument is rather crude and a method for systematically including fluctuations is required. The RG has provided such a method for calculating bulk quantities \cite{PELITI,BPL,DOI,BJH,CT}, with the advantage of clearly identifying universal properties. We now present an overview of the generalisation to a system with a boundary: a case technically more complicated due to the lack of translational invariance. Details of the calculation will be provided elsewhere \cite{RK2}.

The field-theoretic description is obtained by first writing a {\it master equation}. This describes the flow of probability between microstates of the system and is conveniently written in second-quantised form $\partial_t|P\!>=-{\cal H}|P\!>$. The vector $|P\!>$ is the probability-state vector written in a Fock space and acted upon by the evolution operator ${\cal H}$
\begin{eqnarray}
{\cal H}&=& \sum_i\left[D\sum_ja^\dagger_i(a_i-a_j)-\lambda(1-(a^\dagger_i)^2)a_i^2\right] 
\end{eqnarray}
where $a^\dagger,a$ are the usual bosonic operators. The sum $i$ is over all lattice sites, and the sum $j$ is over all of site $i$'s neighbours, with the condition that both sums are restricted to the half-space.

The algebraic description ${\cal H}(a,a^\dagger)$ is mapped onto a continuum path-integral for the action ${\cal S}(\phi,\overline{\phi})$  where the complex fields $\phi$ and $\overline{\phi}$ are analogous to $a$ and $a^\dagger-1$. The action ${\cal S}={\cal S}_D+{\cal S}_R+{\cal S}_{\varrho_0}$ comprises diffusion ${\cal S}_D$, reaction  ${\cal S}_R$ and initial-condition ${\cal S}_{\varrho_0}$ components. The diffusive part provides the propagator for the theory which is Gaussian for the transverse dimensions and has the the following ``mirrored'' form for the $z$-dimension
\begin{eqnarray}
{\cal G}_z(z_f,z_i,t)&=&G(z_f-z_i,t)+G(z_f+z_i,t) 
\end{eqnarray}
where $G(z,t)$ is a Gaussian with a standard deviation of $2Dt$. The other components in the action are
\begin{eqnarray}
{\cal S}_R&=&2\lambda\int_0^tdt\int_{z>0}\!\!\!\!\!d^dr\overline{\phi}\phi^2+\lambda\int_0^tdt\int_{z>0}\!\!\!\!\!d^dr\overline{\phi}^2\phi^2, \label{SR} \\
{\cal S}_{\varrho_0}&=& -\varrho_0\int_0^tdt\int_{z>0}\!\!\!\!\!d^dr\overline{\phi}\delta(t). \nonumber
\end{eqnarray}
The upper critical dimension of the theory is $d_c=2$ and observables were rendered finite by dimensional reguralisation in $d=2-\epsilon$. The propagator is not dressed by the interactions (\ref{SR}), implying the boundary remains effectively reflecting on all scales. In the language of surface critical phenomena, this corresponds to the {\it special transition} \cite{DIEHL} persisting at all orders. This is different from the behaviour frequently seen in equilibrium surface critical phenomena \cite{DD} and in related non-equilibrium systems \cite{DP}. In fact only the reaction rate $\lambda$ is renormalised, with a fixed point structure identical to the bulk case \cite{BPL}. This is understandable as physically the renormalisation of $\lambda$ is connected to the fact that random walks in two dimensions and below are recurrent: a feature unaffected by the presence of a boundary.

To get non-trivial, $z$-dependent results it is clear (from the lack of an excess in the mean-field equation) that the RG improved perturbation expansion must be taken to at least one-loop order. Writing the density as an expansion in $\epsilon=2-d$ and splitting the contribution into an excess $\varrho_E(z,t)$ and a homogeneous, background bulk density $\varrho_B(t)$ 
\begin{eqnarray}
\varrho(z,t)&=&\varrho_B(t)+\varrho_E(z,t), 
\end{eqnarray}
the homogeneous bulk density for $d<2$ is found to be
\begin{eqnarray}
\varrho_B(t)&=& \frac{1}{4\pi\epsilon(Dt)^{d/2}}\left[1+\frac{\epsilon}{4}\left(2\log(8\pi)-5\right)\right]+O(\epsilon).  \label{bulk}
\end{eqnarray}
This is exactly the result found in \cite{BPL} as expected. 

However, a fluctuation-induced density excess is also found, representing the new result from this calculation 
\begin{eqnarray}
\varrho_E(z,t)&=&\frac{1}{8\pi(Dt)^{d/2}}f_2\left(\frac{{z}^2}{2Dt}\right)+O(\epsilon), \label{surf}\\
f_2(\xi^2)&=&\int_0^1 \!ds\int_0^s\!dr \left(\frac{r}{s}\right)^2 \frac{\exp\left(-\frac{\xi^2}{(2-s-r)}\right)} {\left[(s-r)(2-s-r)\right]^{1/2}}. \nonumber
\end{eqnarray} 
The function \cite{int} has the asymptotics $f_2\sim\exp^{-\xi^2/2}/\xi^3$ and is plotted in Fig. 1. A few things should be noted about the form of $\varrho_E$. First, this excess is not localised at the boundary but {\it extends into the system} diffusively, by virtue of its functional dependence of $z^2/Dt$. Also, the excess shares the same universality as the bulk density in that it is independent of the reaction rate $\lambda$ and the initial density $\varrho_0$. Finally, for $d<2$ the amplitude of the excess decays with the same exponent as the bulk. Hence, the ratio of the boundary to the bulk density
\begin{eqnarray}
\frac{\varrho(0,t)}{\varrho_B(t)}&=&1+\epsilon\left(\frac{3}{4}+\frac{\pi}{2}-\frac{3\pi^2}{16}\right)+O(\epsilon^2) \label{ratio}
\end{eqnarray}
is a constant, universal quantity independent of {\it all} system parameters except the dimension. 

The behaviour in two dimensions provides for an interesting result: the one-loop calculation Eq. (\ref{surf}) is the {\it exact} late-time density excess. It is independent from the renormalised reaction rate and therefore represents the universal leading order, with higher-loop corrections decaying as $(t\log t)^{-1}$. The excess given in Eq. (\ref{surf}) gives surprisingly accurate results even for short times, as can be seen in Fig. 1 where the result for the density excess is compared with simulations. 

\begin{figure}
\epsfxsize 8 cm
\epsfysize 7 cm
\epsfbox{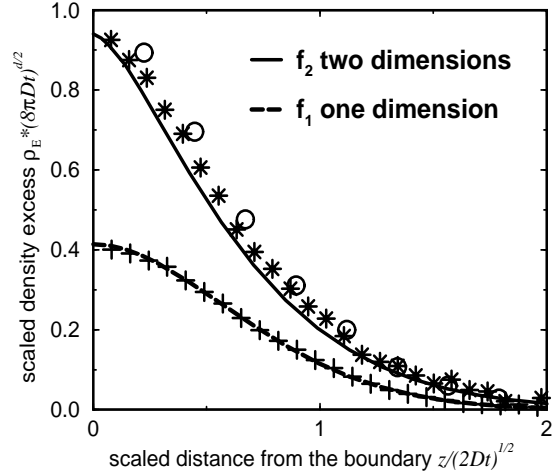}
\caption{The density excess in one and two dimensions, $f_1$ and $f_2$. The RG result $f_2$ is compared with simulations at time $t=10$ ($\circ$) and $t=80$ ($\ast$). The exact result $f_1$ calculated for infinite reaction rate is compared with a simulation for finite reaction rate $\lambda=1/2$ ($+$) demonstrating the universality.}
\vspace{-0.4cm}
\end{figure}

The RG has provided information about the general behaviour of the density excess as a function of dimension. The universal quantities have been identified and the excess density correctly predicted for late times in $d=2$. Unfortunately, the $\epsilon$ expansion gives disappointing results for the amplitude in $d=1$. Motivated by the mapping of the reaction dynamics onto the Ising magnet and the related excitonic coalescence process, we now provide the exact solution in one dimension. We generalise the bulk model described in \cite{LUSH} to include an impenetrable boundary. The on-site reaction rate is chosen to be infinite and hence a site in the system can be occupied by at most one particle. The dynamical rules can be written in terms of the possible evolutions of a pair of neighbouring sites. Denoting a particle on site $k$ by $A_k$ and an empty site by $O_k$, the allowed changes are
\begin{eqnarray}
O_zA_{z+1}&\leftrightarrow&A_zO_{z+1} \mbox{ with a rate D}\label{rates} \\
A_zA_{z+1}&\rightarrow&O_zO_{z+1} \mbox{ with a rate 2D}  \nonumber
\end{eqnarray}
where the site label $z=1,2\cdots$ is restricted to positive integers. The master equation can be written in the language of spin-half operators, and we were able to solve for the density using a similarity transformation found in \cite{SCHUTZ} (details of the calculation will be presented elsewhere \cite{RK2}). The density can be written as a sum of Bessel functions, of which the continuum limit gives the following density excess
\begin{eqnarray}
\varrho_E(z,t)&=&\frac{1}{\sqrt{8\pi Dt}}f_1\left(\frac{z^2}{2Dt}\right), \label{1dexact} \\
f_1(\xi^2)&=&\sqrt{2}\mbox{erfc}\left(\frac{\xi}{\sqrt{2}}\right)\exp\left(\frac{-\xi^2}{2}\right)-\mbox{erfc}\left(\xi\right).\nonumber
\end{eqnarray}
This function has the asymptotics $f_1\sim\exp^{-\xi^2/2}/\xi$ and is plotted in Fig. 1. Again, the bulk component of the density $\varrho_B=1/\sqrt{8\pi Dt}$ was found to be identical to the infinite case \cite{LUSH}. As expected, the form of the solution is in agreement with that predicted from the RG treatment. The universal ratio of the density near the boundary to the bulk density given in Eq. (\ref{ratio}) is found to be exactly $\sqrt{2}$. It should be noted that the one-loop RG prediction for the universal ratio (given in Eq. (\ref{ratio}) with $\epsilon=1$) is numerically $\sim1.47$ for one dimension, and therefore gives quite a fair indication of the exact value. This should be compared with the  poor one-loop predictions for the amplitudes of the density itself in one dimension.

The universality (independence from the reaction rate $\lambda$) predicted by the RG treatment, can be seen by comparing this exact result for $\lambda=\infty$ with data from a simulation of a system with finite reaction rate (Fig. 1).

The result, Eq. (\ref{1dexact}), gives the time-dependent probability density of domain walls in the coarsening, one-dimensional critical Ising model near a very strong magnetic field, i.e. a fixed spin. The magnitude of the coarse-grained magnetisation is a function of the local density of domain walls - the fewer the domain walls the higher the magnetisation. Hence, the result implies that the absolute value of the magnetisation measured near a fixed spin is lower than in the bulk of the system. It would be very interesting to see if this dynamic effect is seen in other magnetic systems or in higher dimensions.

In summary, results have been presented from an analysis of a reaction-diffusion process near an impenetrable boundary. The mean-field equation was shown to predict a flat density profile. However, it was demonstrated that in two dimensions and below, a density excess develops at the boundary. In one dimension it was found that the density excess is as significant a contribution as the bulk density with both decaying as $\sim t^{-1/2}$. In two dimensions the excess was found to be marginally subdominant, decaying as $t^{-1}$. Both these density excesses shares the same universality as the bulk density, in that they are independent of the reaction rate and the initial density. Moreover, a higher degree of universality was found in the ratio between the boundary and bulk densities: a quantity depending only on the dimension of space. The functional forms for the density excess were obtained for both one and two dimensions and a link was made to the coarsening dynamics of the one-dimensional Ising model. Most importantly, it was shown that the excess is not localised at the boundary but extends diffusively throughout the bulk of the system. It should be noted that above the critical dimensions there is a sub-dominant density excess. However, these effects are transient and quickly decay to yield the mean-field result, as predicted.

A quantity much studied recently and related to domain coarsening dynamics is the {\it persistence} exponent. This exponent describes the time dependence of the distribution of sites not yet visited by a domain wall. In the homogeneous bulk case, persistence has been examined by the RG \cite{CARDY,KLAUS} and an exact result found in one dimension \cite{DERR}. In light of result (\ref{1dexact}) and the fact that the excess penetrates into the system, it would be worth examining the behaviour of the persistence in the presence of a fixed spin.

Finally, it is interesting to consider the difference between the behaviour of diffusing and gas-phase ballistic reaction kinetics near a boundary.  A simple realisation of the ballistic $A\!+\!A\!\!\rightarrow\!\!O$ system \cite{EF} shows the same bulk density-decay exponent in one dimension as the diffusive case. However, contrary to the diffusive case, it can be shown that for ballistic reactions with an (elastic) impenetrable boundary there is a {\it lower} density of reactants near the wall \cite{RK2}. This is understandable, because in late times the remaining particles tend to congregate in groups moving in the same direction. When such a group hits the elastic wall it mostly annihilates within itself, leaving few particles to return. In the context of surface critical phenomena this corresponds to the impenetrable boundary behaving as an effectively {\it absorbing} boundary on long time scales, in contrast to the diffusive case.\\

{\it Acknowledgments.}
We would like to thank David Mukamel and Gunter Sch\"utz for useful discussions. Fabian Essler, Martin Evans and Klaus Oerding are also thanked for their careful reading of the manuscript. The authors acknowledge support from the Israeli Science Foundation.
\vspace{-0.6cm}


\begin{thebibliography}{100}
\bibitem{int}
We were unable to perform the full integration except at the point $z=0$: $f_2(0)=3/2+\pi-3\pi^2/8$.
\end{thebibliography}
\end{document}